\def\be{\begin{equation}}
\def\ee{\end{equation}}
\def\ba{\begin{array}}
\def\ea{\end{array}}

\documentclass[aps,amsmath,amssymb,amsfonts,showpacs]{revtex4}
\usepackage{graphicx}
\usepackage{epstopdf}
\def\qed{\leavevmode\unskip\penalty9999 \hbox{}\nobreak\hfill
     \quad\hbox{\leavevmode  \hbox to.77778em{%
               \hfil\vrule   \vbox to.675em%
               {\hrule width.6em\vfil\hrule}\vrule\hfil}}
     \par\vskip3pt}

\newtheorem{theorem}{Theorem}
\newtheorem{corollary}{Corollary}
\newtheorem{lemma}{Lemma}
\newcommand{\ket}[1]{|#1\rangle}
\newcommand{\bra}[1]{\langle#1|}

\begin{document}

\title{ Monogamy relations and upper bounds for the generalized $W$-class states using R\'{e}nyi-$\alpha$ entropy}

\author{Yanying Liang$^1$, Zhu-Jun Zheng$^1$, Chuan-Jie Zhu$^2$}

\affiliation{$^1$School of Mathematics, South China University of Technology, Guangzhou 510641, China\\
$^2$College of Mathematics and Physics Science, 
Hunan University of Arts and Science, Changde 415000, P.R. China\\
 $^2$Department of Physics, Renmin University of China,   Beijing 100872,  China}

\begin{abstract}

We investigate monogamy relations and upper bounds for generalized $W$-class states related to the R\'{e}nyi-$\alpha$ entropy. First, we present an analytical formula on R\'{e}nyi-$\alpha$ entanglement (R$\alpha$E) and R\'{e}nyi-$\alpha$ entanglement of assistance (REoA) of a reduced density matrix for a generalized $W$-class states. According to the analytical formula, we show monogamy and polygamy relations for generalized $W$-class states in terms of R$\alpha$E and REoA. Then we give the upper bounds for generalized $W$-class states in terms of R$\alpha$E. Next, we provide tighter monogamy relations for generalized $W$-class states in terms of concurrence and convex-roof extended negativity and obtain the monogamy relations for R$\alpha$E by the analytical expression between R$\alpha$E and concurrence. Finally, we apply our results into quantum games and present a new bound of the nonclassicality of quantum games restricting to generalized $W$-class states.

\end{abstract}


\maketitle

\section{Introduction}
In a multipartite quantum system, if a pair of parties share maximal entanglement, according to its restricted sharability , they can share neither entanglement \cite{091,092} and nor classical correlations \cite{093} with the rest. This is known as monogamy of entanglement (MoE) \cite{094}. Since MoE can quantify how much information an eavesdropper could  potentially obtain about the secret key to be extracted,  MoE is a key ingredient to make quantum cryptography secure \cite{20205,20206}.

The first mathematical characterization of MoE \cite{091} was shown by Coffman, Kundu and Wootters using squared concurrence \cite{Wootters1998}. It is known as CKW-inequality. Later the CKW-type inequality was shown for arbitrary multiqubit systems \cite{092}. The MoE of squared concurrence can be used to characterize the entanglement structure in multipartite quantum systems and detect the existence of multiqubit entanglement in dynamical procedures \cite{YKB1,YKB2,Ou,Jung}. Furthermore,
there are also many works devoted to the topic of entanglement monogamy \cite{yyl2019,yyl2020,YCS,DY}.

However, monogamy relations using concurrence is known to fail in the generalization of CKW inequaity for higher dimensional quantum systems \cite{yym18}. In three-qubit quantum system, it is well-known that there exists two inequivalent classes of genuine tripartite entangled states. The first is the Greenberger-Horne-Zeilinger (GHZ) class~\cite{GHZ}, the other one is the W-class~\cite{DVC}. The conversion of the states in a same class can be achieved by local operation and classical communication with non-zero probability.  CKW inequality is saturated by W-class states and it becomes the most strict inequality with the states in GHZ class \cite{KS}. The saturation of the inequality implies  a genuine tripartite entanglement could have a complete characterization by the bipartite ones inside it. So in this paper, we are interested in the monogamy relations of the $n$-qubit generalized $W$-class states proposed in \cite{KS}.

Moreover in \cite{KS}, Kim and Sanders showed that the entanglement of the $n$-qubit generalized $W$-class states is fully characterized by their partial entanglements using squared concurrence. In 2014, Kim considered a large class of multi-qubit generalized W-class states, and analytically showed  the strong monogamy inequality
of multi-qubit entanglement is saturated by this class of states \cite{Kim14}.
In 2015, Choi and Kim  provided an analytical proof that strong monogamy inequality of squared convex-roof extended negativity is saturated by a large class of multi-qudit states; a superposition of multi-qudit generalized W-class states and vacuums \cite{Kim15}. In 2016, Kim showed some useful properties for a large class of multi-qudit mixed state that are in a partially coherent superposition of a generalized $W$-class state and the vacuum \cite{Kim16}.

R\'{e}nyi-$\alpha$ entanglement (R$\alpha$E) \cite{hor96pla} is a well-defined entanglement measure
which is the generalization of entanglement of formation (EOF) and has the merits for characterizing
quantum phases with differing computational power \cite{jcui12nc}, ground state properties in many-body
systems \cite{fran14prx}, and topologically ordered states \cite{fla09prl,hal13prl}. Although EoF
is known to fail for usual CKW-type characterization of MoE,
R\'enyi-$\alpha$ entropy can still be shown to have CKW-type
monogamy inequality for all case of $\alpha$ if it exceeds a certain
threshold \cite{JSK4}. Kim proved monogamy of entanglement in multi-qubit systems for $\alpha
\geq 2$ using R\'enyi-$\alpha$ entanglement to quantify bipartite entanglement \cite{JSK4}. Wei \emph{et al} presented the squared R\'{e}nyi-$\alpha$ entanglement (SR$\alpha$E) obeys a general monogamy inequality in an arbitrary $N$-qubit mixed state. In 2016, Wei \emph{et al} presented lower and upper bounds for R\'enyi-$\alpha$ entanglement \cite{WS1}.

Apart from the entertainment value, games among multiplayers often provide an intuitive way to understand complex problems. In Ref. \cite{quantumgame1}, Marco \emph{et al} investigated the probability that both players in a quantum game simultaneously succeed in guessing the outcome correctly. Their results implies the optimal guessing probability can be achieved without the use of entanglement. In Ref. \cite{2020,quantumgame}, the authors presented bounds on the difference between multiplayer quantum games and classical games using the monogamy of Tsallis-$q$ entropy and squashed entanglement, respectively.

This paper is organized as follows. In Sec.\ref{2}, we give the preliminary knowledge needed in this paper. In Sec.\ref{A}, we show monogamy and polygamy relations for $n$-qubit GW states using R$\alpha$E and REoA. We generalize these relations into $\mu$-th power of R$\alpha$E when $\mu \geq 2$ and REoA when $0 <\mu \leq 1$. In Sec.\ref{B}, we present the upper bounds for $n$-qubit GW states in terms of R$\alpha$E. In Sec.\ref{C}, we obtain tighter monogamy relations using concurrence and CREN. We also analysis the general monogamy relations for $n$-qubit GW states using R$\alpha$E according to the analytical expression between the R$\alpha$E and concurrence. In Sec.\ref{4}, we provide a new bound on the difference between multiplayer quantum games and classical games restricting to the $n$-qubit GW states using monogamy of R$\alpha$E. In Sec.\ref{5}, we end with a conclusion.
\section{Preliminary Knowledge}\label{2}

For a bipartite pure state $\ket{\psi}_{AB}=\sum_i\sqrt{\lambda_i}\ket{ii},$  the concurrence  $C(|\psi\rangle_{AB})$ is defined as \cite{PR}
\begin{equation}\label{cpure}
\mathcal{C}(|\psi\rangle_{AB}) = \sqrt{2[1-Tr(\rho_{A}^{2})]},
\end{equation}
where $\rho_{A} = Tr_{B}(|\psi\rangle_{AB}\langle\psi|)$ (and analogously for $\rho_{B}$).

For any mixed state $\rho_{AB}, $ the concurrence is given visa the so-called convex roof extension
\begin{equation}
\mathcal{C}(\rho_{AB}) = \min_{\{p_{i}, |\psi_{i}\rangle\}}\sum_{i}p_{i}\mathcal{C}(|\psi_{i}\rangle),
\end{equation}
where the minimum is taken over all possible pure decompositions of $\rho _{AB}= \sum_{i} {p_i | \psi _i \rangle _{AB} \langle \psi _i |}$.

As the duality of concurrence, the concurrence of assistance (CoA) of any mixed state $\rho_{AB}$ is defined as \cite{CSY}
\begin{equation}
\mathcal{C}_{a}(\rho_{AB}) = \max_{\{p_{i}, |\psi_{i}\rangle\}}\sum_{i}p_{i}\mathcal{C}(|\psi_{i}\rangle),
\end{equation}
where the maximum is taken over all possible pure state decompositions $\{p_{i}, |\psi_{i}\rangle\}$ of $\rho_{AB}$.

A well-known quantification of bipartite entanglement is negativity \cite{GV}, which is based on the positive partial transposition (PPT) criterion.
For a bipartite state $\rho_{AB}$ in a $d \otimes d^{'} (d\leq d^{'})$ quantum system,  its negativity is defined as
\begin{equation}
\mathcal{N}(\rho_{AB}) = \|\rho_{AB}^{T_A}\|-1,
\end{equation}
where $\rho_{AB}^{T_A}$ is the partial transpose with respect to the subsystem $A$ and $\|X\|$ denotes the trace norm of $X, $  $\|X\| = Tr \sqrt{XX^{\dagger}}.$

To overcome the lack of separability criterion, one modification of negativity is convex-roof extended negativity (CREN), which gives a perfect discrimination of PPT bound entangled states and separable states in any bipartite quantum system \cite{yym15}. For a bipartite mixed state $\rho_{AB}, $ CREN is defined as
\begin{equation}
\widetilde{\mathcal{N}}(\rho_{AB}) = \min_{\{p_{i}, |\psi_{i}\rangle\}}\sum_{i}p_{i}\mathcal{N}(|\psi_{i}\rangle),
\end{equation}
where the minimum is taken over all possible pure state decompositions $\{p_{i}, |\psi_{i}\rangle\}$ of $\rho_{AB}$.

Similar to the duality between concurrence and CoA,  we can also define a dual to CREN,  namely CRENoA,  by taking the maximum value of average negativity over all
possible pure state decomposition \cite{yym18},  i.e.
\begin{equation}
\widetilde{\mathcal{N}}_{a}(\rho_{AB}) = \max_{\{p_{i}, |\psi_{i}\rangle\}}\sum_{i}p_{i}\mathcal{N}(|\psi_{i}\rangle),
\end{equation}
where the maximum is taken over all possible pure state decompositions $\{p_{i}, |\psi_{i}\rangle\}$ of $\rho_{AB}$.

Another well-known quantification of bipartite entanglement is R\'{e}nyi-$\alpha$ entanglement (R$\alpha$E) \cite{JSK4}.
For a bipartite pure state  $\ket{\psi}_{AB}=\sum_i\sqrt{\lambda_i}\ket{ii},$  the R$\alpha$E is defined as
\begin{equation}\label{epure}
E_{\alpha}(| \psi \rangle _{AB})= S_\alpha(\rho_A) = \frac{1}{1-\alpha}\log _2(\mbox{tr}\rho _A^\alpha),
\end{equation}
where the R\'{e}nyi-$\alpha$ entropy is $S_\alpha(\rho_A)=[\log _2(\sum_i\lambda_i^{\alpha})]/(1-\alpha)$ with
$\alpha$ being a nonnegative real number and $\lambda_i$ being the eigenvalue of reduced density matrix
$\rho_A$. The R\'{e}nyi-$\alpha$ entropy $S_\alpha \left( \rho  \right)$ converges to the von Neumann
entropy when the order $\alpha$ tends to 1.

For a bipartite mixed state $\rho _{AB}$, the R$\alpha$E
is defined via the convex-roof extension
\begin{eqnarray}\label{q4}
E_\alpha(\rho _{AB})=\min \sum_i p_i E_\alpha(|\psi _i\rangle_{AB}),
\end{eqnarray}
where the minimum is taken over all possible pure state decompositions of $\rho _{AB}= \sum_{i} {p_i | \psi _i \rangle _{AB} \langle \psi _i |}$.

As a dual concept to R\'{e}nyi-$\alpha$ entanglement, we define the R\'{e}nyi-$\alpha$ entanglement of assistance (REoA) as \cite{JSK4}
\begin{eqnarray}\label{q7}
E_\alpha ^a \left( {\rho _{AB} } \right) = \max \sum_i {p_i } E_\alpha  \left( {\left| {\psi _i } \right\rangle _{AB} } \right),
\end{eqnarray}
where the maximum is taken over all possible pure state decompositions of $\rho _{AB}= \sum_{i} {p_i | \psi _i \rangle _{AB} \langle \psi _i |}$.

In particular, for any two-qubit pure state with its Schmidt decomposition
$\ket{\psi}_{AB}=\sqrt{\lambda_{0}}\ket{00}_{AB}+\sqrt{\lambda_{1}}\ket{11}_{AB}$,
then we have $  C^2(\ket{\psi}_{AB})=4\lambda_0\lambda_1,$ and
\begin{eqnarray}
E_{\alpha}\left(\ket{\psi}_{AB} \right)&=&S_{\alpha}(\rho_A)\nonumber\\
&=&\frac{1}{1-\alpha}\log\left(\lambda_0^{\alpha}+\lambda_1^{\alpha} \right).
\label{RM pure}
\end{eqnarray}
from the above equalities $(\ref{cpure})$ and $(\ref{epure})$.
Then we have an analytical expression between the R\'{e}nyi-$\alpha$ entanglement and concurrence for any two-qubit pure state \cite{JSK4,WS,YXW}
\begin{eqnarray}\label{q12}
E_\alpha  \left( {\ket{\psi}_{AB} } \right) = f_\alpha  \left[ {C^2 \left( {\ket{\psi}_{AB} } \right)} \right],
\end{eqnarray}
where the order $\alpha\geq(\sqrt 7  - 1)/2$ and the function $f_\alpha \left( x \right)$ has the form
\begin{equation}\label{falpha}
f_\alpha \! \left( x \right)\!= \!\frac{1}{{1 - \alpha }}\!\log _2 \!\left[ {\left( {\frac{{1 \!-\!
\sqrt {1 - x} }}{2}} \right)^\alpha  \!\!\!\! +\! \left( {\frac{{1 \!+\! \sqrt {1 - x} }}{2}}
\right)^\alpha  } \right].
\end{equation}

Now we present several lemmas on the properties of the function $f_\alpha \left( x \right)$ in equality $(\ref{falpha})$.

\begin{lemma} \cite{WS}
	The function $f_\alpha^2(x)$ with $\alpha\geq(\sqrt 7  - 1)/2$ is monotonically increasing and convex.
\end{lemma}
\begin{lemma} \cite{WS}
	The function $f_\alpha(x)$ is monotonically increasing and concave for $\alpha\in[(\sqrt7-1)/2,(\sqrt{13}-1)/2]$.
\end{lemma}
Set $y=x^2$, $g_\alpha(y)=f_\alpha(x^2)$.
\begin{lemma} \cite{JSK4,YXW}
	The function $g_\alpha(y)$ is a monotonically increasing and convex function for $0\leq y \leq1,$ and $\alpha \geq(\sqrt7-1)/2$.
\end{lemma}

Next let us recall a class of multi-qubit generalized W-class (GW) states~\cite{KS,Kim14,Kim15}
\begin{align}
\ket{\psi}_{A_1 A_2 ... A_n}=&
a_1 \ket{10\cdots0}+a_2 \ket{01\cdots0}+...+a_n \ket{00\cdots1}
\label{supWV}
\end{align}

\begin{align}
\left|W_n^d \right\rangle_{A_1\cdots A_n}=\sum_{i=1}^{d-1}(&a_{1i}{\ket {i0\cdots 0}} +a_{2i}{\ket {0i\cdots 0}}+\cdots +a_{ni}{\ket {00\cdots 0i}}),
\label{generalWstate}
\end{align}
with the normalization condition $\sum_{i=1}^{n}|a_j|^2 =1$ and  $\sum_{s=1}^{n}\sum_{i=1}^{d-1}|a_{si}|^2=1$ respectively.

The state in Eq.~(\ref{generalWstate}) is a coherent superposition of all $n$-qudit
product states with Hamming weight one. Eq.~(\ref{generalWstate}) includes $n$-qubit W-class states in Eq.~(\ref{supWV}) as a special case when $d=2$.

Next thing we need to do is to present some lemmas for the multi-qubit generalized W-class states which are useful in the proof of our main results.
\begin{lemma} \cite{Kim15}
Let $\ket{\psi}_{A_1\cdots A_n}$ be a $n$-qudit pure state in a superposition of a $n$-qudit generalized W-class state
in Eq.~(\ref{generalWstate}) and vacuum, that is,
\begin{equation}
\ket{\psi}_{A_1A_2\cdots A_n}=\sqrt{p}\left|W_n^d \right\rangle_{A_1\cdots A_n}+\sqrt{1-p}\ket{0\cdots 0}_{A_1\cdots A_n}
\label{WV2}
\end{equation}
for $0\leq p \leq 1$. Let $\rho_{A_1A_{j_1}\cdots A_{j_{m-1}}}$ be a reduced density matrix of $\ket{\psi}_{A_1\cdots A_n}$
onto $m$-qudit subsystems $A_1A_{j_1}\cdots A_{j_{m-1}}$ with $2 \leq m \leq  n-1$.
For any pure state decomposition of $\rho_{A_1A_{j_1}\cdots A_{j_{m-1}}}$ such that
\begin{align}
\rho_{A_1A_{j_1}\cdots A_{j_{m-1}}}=\sum_{k}q_k\ket{\phi_k}_{A_1A_{j_1}\cdots A_{j_{m-1}}}\bra{\phi_k},
\label{rhoa1aj1ajm-1}
\end{align}
$\ket{\phi_k}_{A_1A_{j_1}\cdots A_{j_{m-1}}}$ is a superposition of a $m$-qudit generalized W-class state and vacuum.
\label{lemma: reduced}
\end{lemma}

\begin{lemma}\cite{KS}
 For any $n$-qubit W-class states $\ket{\psi}_{A_1\cdots A_n}$ and a partition $P=\{P_1,\ldots,P_m \}$
of the set of subsystems $S=\{A, B_1,\ldots,B_{n-1} \}$,$m\leq n$
\begin{equation}\label{lemma5-1}
\mathcal{C}_{P_s(P_1\cdots\widehat{P}_s\cdots P_m)}^2 = \sum_{k \neq s}\mathcal{C}_{P_s P_k}^2 = \sum_{k \neq s}(\mathcal{C}_{P_s P_k}^a)^2,
\end{equation}
and
\begin{equation}\label{lemma5-2}
\mathcal{C}_{P_s P_k}=(\mathcal{C}_{P_s P_k}^a),
\end{equation}
for all $k \neq s$ and $(P_1\cdots\widehat{P}_s\cdots P_m)=(P_1\cdots{P}_s\cdots P_m)-(P_s)$.
\end{lemma}
\begin{lemma} \cite{Kim16}
Let $\ket{\psi}_{A, B_1,\cdots,B_{n-1}}$ be a $n$-qudit pure state in a superposition of a $n$-qudit generalized W-class state in Eq.~(\ref{generalWstate}) and vacuum, then for any partition $P=\{P_1,\ldots,P_m \}$
of the set of subsystems $S=\{A, B_1,\ldots,B_{n-1} \}$,$m\leq n$, the state $\ket{\psi}_{P_1,\cdots,P_{m}}$ is also a
superposition of a $n$-qudit generalized W-class state in Eq.~(\ref{generalWstate}) and vacuum. Here $P_s \cap P_t = \emptyset~\mathrm{for}~s\neq t,$ and $\bigcup_s P_s = S$.
\end{lemma}
At last, we have one more lemma which is used in the last part of our main results.
\begin{lemma}
	For real numbers $t\in[0,1]$ , $x\geq k \geq1$, we have
	\begin{align}\label{lem}
	(1+x)^t\geq1+\left(\frac{(1+k)^x-1}{k^x}\right)x^t.
	\end{align}
\end{lemma}
[Proof]
Consider the function $f_t(x)=\frac{(1+x)^t-1}{x^t}$. Since
$$\frac{d f_{t}(x)}{d x}=tx^{-(t+1)}[1-(1+x)^{t-1}]\geq0,$$
for $t\in[0,1]$ and $x\geq1$.

In other words, the  function $f_t(x)$ is an increasing function with $x\geq1$.
Since $x\geq k \geq1$, then $f_t(x)\geq f_t(k)$.
\qed

\section{Main Results}
In this section, we give the main results of this paper.
In Sec.\ref{A}, we show monogamy and polygamy relations for R$\alpha$E and REoA of GW states, and generalize them into the $\mu$-th power of R$\alpha$E for $\mu \geq2$ and $\mu$-th power of REoA for $0< \mu \leq 1$.
In Sec.\ref{B}, we investigate the upper bounds for GW states using R$\alpha$E.
In Sec.\ref{C}, we present tighter monogamy relations in terms of concurrence and CREN. We also get the general monogamy relations for R$\alpha$E using the analytical expression between the R$\alpha$E and concurrence.

\subsection{Monogamy and polygamy relations using R\'{e}nyi entropy for generalized W-class states}\label{A}
For a pure GW state, we have the following theorem.
\begin{theorem}\label{Ralpha1}
	Assume $\rho_{A_{j_1}\cdots A_{j_m}}$ is a reduced density matrix of a pure GW state, then we have
	\begin{align}
	E_\alpha(\rho_{A_{j_1}|A_{j_2}\cdots A_{j_m}})=f_\alpha(C^2(\rho_{A_{j_1}|A_{j_2}\cdots A_{j_m}})),
	\end{align}
	when $\alpha\geq(\sqrt 7  - 1)/2$.
\end{theorem}

The proof is similar to the proof of Theorem 1 in Ref. \cite{JSK4} and Lemma 1 is also needed in the proof. Next we give an analytic formula of REoA for a GW state.

\begin{theorem}\label{Ralpha2}
	Assume $\rho_{A_{j_1}\cdots A_{j_m}}$ is a reduced density matrix of a pure GW state, then we have
	\begin{align}
	E_\alpha^a(\rho_{A_{j_1}|A_{j_2}\cdots A_{j_m}})=f_\alpha(C^2(\rho_{A_{j_1}|A_{j_2}\cdots A_{j_m}})),
	\end{align}
	when $\alpha\in[(\sqrt7-1)/2,(\sqrt{13}-1)/2]$.
\end{theorem}
[Proof]
For convenience, we denote $\rho_{A_{j_1}|A_{j_2}\cdots A_{j_m}}$ as $\rho_{AB}$. From \cite{Kim14,KS,yyl2017}, if $\rho_{AB}$ is a reduced density matrix of a pure GW state, then we have $C(\rho_{AB})=C_a(\rho_{AB})$.
So it is enough for us to show $E_\alpha^a(\rho_{AB})= f_\alpha(C_a^2(\rho_{AB}))$.
First we prove $E_\alpha^a(\rho_{AB})\leq f_\alpha(C_a^2(\rho_{AB}))$. Assume $\{p_{i}, |\psi_{i}\rangle\}$ is the optimal decomposition for REoA of $\rho_{AB}$, then we have
\begin{align}\label{da}
	E_\alpha^a(\rho_{AB})=&\sum_i p_i E_\alpha(\ket{\psi_i}_{AB})\nonumber\\
	=&\sum_i p_i f_\alpha(C^2(\ket{\psi_i}_{AB}))\nonumber\\
	\leq & f_\alpha(\sum_i p_iC^2(\ket{\psi_i}_{AB}))\nonumber\\
\leq & f_\alpha(C_a^2(\rho_{AB})),
\end{align}
where the first inequality is due to the concave property of $f_\alpha(x)$ for $\alpha\in[(\sqrt7-1)/2,(\sqrt{13}-1)/2]$ in Lemma 2 and the second inequality is due to the definition of $C_a^2(\rho_{AB})$ and the increasing property of $f_\alpha(x)$ in Lemma 2.

Next we show $E_\alpha^a(\rho_{AB})\geq f_\alpha(C_a^2(\rho_{AB}))$. Set $y=x^2, g_\alpha(y)=f_\alpha(x^2).$
Assume $\{r_{k}, |\theta_{k}\rangle\}$ is the optimal decomposition for $C_a(\rho_{AB})$.
Then we have
\begin{align}\label{xiao}
	g_\alpha(C_a(\rho_{AB}))= & g_\alpha(\sum_k r_k C(\ket{\theta_k}_{AB}))\nonumber\\
\leq &\sum_k r_k g_\alpha(C(\ket{\theta_k}_{AB})))\nonumber\\
	= & \sum_k r_k E_\alpha(\ket{\theta_k}_{AB})\nonumber\\
\leq & E_\alpha^a(\rho_{AB}),
\end{align}
where in the first inequality we have used the convex property of $g_\alpha(y)$ for $\alpha \geq(\sqrt7-1)/2$ in Lemma 3. The second inequality is due to the definition of $E_\alpha^a(\rho_{AB})$. Since $y=x^2$ and let $x=C_a(\rho_{AB})$, then we have $E_\alpha^a(\rho_{AB})\geq f_\alpha(C_a^2(\rho_{AB}))$.

Thus combining (\ref{da}) and (\ref{xiao}), we have
$E_\alpha^a(\rho_{AB})= f_\alpha(C_a^2(\rho_{AB}))=f_\alpha(C^2(\rho_{AB}))$ which completes the proof.
\qed

According to Theorem 1 and Theorem 2, we have the following Theorem 3.
\begin{theorem}\label{Ralpha3}
Assume $\rho_{A_{j_1}\cdots A_{j_m}}$ is a reduced density matrix of a pure GW state, then we have
\begin{align}
E_\alpha^a(\rho_{A_{j_1}|A_{j_2}\cdots A_{j_m}})=E_\alpha(\rho_{A_{j_1}|A_{j_2}\cdots A_{j_m}})=f_\alpha(C^2(\rho_{A_{j_1}|A_{j_2}\cdots A_{j_m}})),
\end{align}
when $\alpha\in[(\sqrt7-1)/2,(\sqrt{13}-1)/2]$.
\end{theorem}

Now we begin to investigate the monogamy relation using R\'{e}nyi entropy for generalized W-class states.
\begin{theorem}\label{monogamy1}
	Assume $\rho_{A_{j_1}A_{j_2}\cdots A_{j_m}}$ is the reduced density matrix of a GW state $\ket{\psi}_{A_1\cdots A_n},$ and here we denote $\{P_1,P_2,\cdots,P_k\}$ is a partition of the set $\{A_{j_1},A_{j_2},\cdots,A_{j_m}\},$ when $\alpha\geq(\sqrt 7  - 1)/2$, we have the following monogamy inequality
	\begin{align}
	E_\alpha^2(\rho_{P_1|P_2\cdots P_k})\geq \sum_{i=2}^{k} E_\alpha^2(\rho_{P_1P_i}).
	\end{align}
\end{theorem}
[proof]
For $\alpha\geq(\sqrt 7  - 1)/2$, we have
\begin{align}
	E_\alpha^2(\rho_{P_1|P_2\cdots P_k})=&f_\alpha(C^2(\rho_{P_1|P_2\cdots P_k}))\nonumber\\
	=& f_\alpha(\sum_{i=2}^{k}C^2(\rho_{P_1P_i}))\nonumber\\
	\geq & \sum_{i=2}^{k}f_\alpha(C^2(\rho_{P_1P_i}))\nonumber\\
= & \sum_{i=2}^{k} E_\alpha^2(\rho_{P_1P_i}),
\end{align}
where in the second equality we use Lemma 5 and the inequality is due to Lemma 1.
\qed

Naturally we want to generalize Theorem~\ref{monogamy1} into the $\mu$-th power of R$\alpha$E for GW states when $\mu\geq2$. We find that
when $k=3$, we can always get $ E_\alpha^2(\rho_{P_1P_3})\leq  E_\alpha^2(\rho_{P_1P_2})$ through designing the partition $\{P_1,P_2,P_3\}$, then we get
\begin{eqnarray*}
  E_\alpha^{\mu}(\rho_{P_1|P_2P_3})
  && \geq (E_\alpha^2(\rho_{P_1P_2})+E_\alpha^2(\rho_{P_1P_3}))^{\frac{\mu}{2}}\\
  &&=E_\alpha^\mu(\rho_{P_1|P_2})\left(1+\frac{E_\alpha^2(\rho_{P_1P_3})}{E_\alpha^2(\rho_{P_1P_2})}\right)^{\frac{\mu}{2}}\\
  && \geq
  E_\alpha^\mu(\rho_{P_1P_2})+E_\alpha^\mu(\rho_{P_1P_3})
\end{eqnarray*}
where in the first inequality we use Theorem~\ref{monogamy1}. The second inequality is obtained by $(1+t)^x\geq 1+t^x$ for any real number $x$ and $t$, $0 \leq t  \leq 1$, $x\in [1, \infty]$.

Therefore we can have the following corollary by the way of  use this operation repeatedly.
\begin{corollary}\label{monogamy2}
	Assume $\rho_{A_{j_1}A_{j_2}\cdots A_{j_m}}$ is the reduced density matrix of a GW state $\ket{\psi}_{A_1\cdots A_n},$ and here we denote $\{P_1,P_2,\cdots,P_k\}$ is a partition of the set $\{A_{j_1},A_{j_2},\cdots,A_{j_m}\},$ when $\alpha\geq(\sqrt 7  - 1)/2$, we have the following monogamy inequality,
	\begin{align}
	E_\alpha^\mu(\rho_{P_1|P_2\cdots P_k})\geq \sum_{i=2}^{k} E_\alpha^\mu(\rho_{P_1P_i}).
	\end{align}
for $\mu\geq2$.
\end{corollary}

As a duality of monogamy relations, polygamy relations using REoA for GW states can also be developed.
\begin{theorem}\label{polygamy1}
	Assume $\rho_{A_{j_1}A_{j_2}\cdots A_{j_m}}$ is the reduced density matrix of a GW state $\ket{\psi}_{A_1\cdots A_n},$ and here we denote $\{P_1,P_2,\cdots,P_k\}$ is a partition of the set $\{A_{j_1},A_{j_2},\cdots,A_{j_m}\},$ when $\alpha\in[(\sqrt7-1)/2,(\sqrt{13}-1)/2]$, we have the following polygamy inequality,
	\begin{align}
	E_\alpha^a(\rho_{P_1|P_2\cdots P_k})\leq \sum_{i=2}^{k} E_\alpha^(\rho_{P_1P_i}).
	\end{align}
\end{theorem}
[proof]
From Theorem~\ref{Ralpha2}, we have
\begin{align}
	E_\alpha^a(\rho_{P_1|P_2\cdots P_k})=&f_\alpha(C^2(\rho_{P_1|P_2\cdots P_k}))\nonumber\\
	=&f_\alpha(\sum_{i=2}^{k} C^2(\rho_{P_1P_i}))\nonumber\\
	\leq & \sum_{i=2}^{k} f_\alpha(C^2(\rho_{P_1P_i}))\nonumber\\
	=& \sum_{i=2}^{k} E_\alpha(\rho_{P_1P_i}),
\end{align}
where the inequality is due to Lemma 2.
\qed

When $0<\mu \leq1$, using $(1+t)^x\geq 1+t^x$ with $0\leq t \leq1$, $x\in[0,1]$ and similar method in Corollary~\ref{monogamy2}, we have the following corollary.
\begin{corollary}\label{polygamy2}
	Assume $\rho_{A_{j_1}A_{j_2}\cdots A_{j_m}}$ is the reduced density matrix of a GW state $\ket{\psi}_{A_1\cdots A_n},$ and here we denote $\{P_1,P_2,\cdots,P_k\}$ is a partition of the set $\{A_{j_1},A_{j_2},\cdots,A_{j_m}\},$ when $\alpha\in[(\sqrt7-1)/2,(\sqrt{13}-1)/2]$£¬ we have the following polygamy inequality,
	\begin{align}
	(E_\alpha^a)^\mu(\rho_{P_1|P_2\cdots P_k})\leq \sum_{i=2}^{k} E_\alpha^\mu(\rho_{P_1P_i}).
	\end{align}
for $0< \mu \leq1$.
\end{corollary}

The method which has been used so far  can be generalized to investigate monogamy and polygamy inequalities using other entanglement measures for GW states, such as Tsallis $q$ entropy \cite{2020} and unified entropy~\cite{31,32}.

As an example, we consider the 4-qubit generalized W-class state
\begin{align}
&\ket{\psi}_{A_1A_2A_3A_4}
=0.3\ket{0001}+0.4\ket{0010}+{0.5}\ket{0100}+\sqrt{0.5}\ket{1000},
\end{align}

Here we choose $\rho_{A_1A_2A_3}$ is the reduced density matrix of $\ket{\psi}_{A_1A_2A_3A_4}$, $P_1=A_1, P_2=A_2, P_3=A_3$. Then we have
\begin{align}
\rho_{A_1A_2A_3}=&0.09\ket{000}\bra{000}+\ket{\phi}\bra{\phi}
\end{align}
where $\ket{\phi}=0.4\ket{001}+0.5\ket{010}+\sqrt{0.5}\ket{100}$.
After calculation, we get $C(\rho_{P_1P_2})=\frac{\sqrt{2}}{2}$, $C(\rho_{P_1P_3})=\frac{2\sqrt{2}}{5}$.
Then from Theorem \ref{Ralpha1}, we have
\begin{align}
E_\alpha(\rho_{P_1P_2})=&f_\alpha\left((\frac{\sqrt{2}}{2})^2\right),\nonumber\\
E_\alpha(\rho_{P_1P_3})=&f_\alpha\left((\frac{2\sqrt{2}}{5})^2\right),\nonumber
\end{align}

Combining Theorem \ref{Ralpha3}, Theorem \ref{monogamy1}, and Theorem \ref{polygamy1} , we have
\begin{align}
\sqrt{E_\alpha^2(\rho_{P_1P_2})+E_\alpha^2(\rho_{P_1P_3})}\leq E_\alpha(\rho_{P_1|P_2P_3})\leq E_\alpha(\rho_{P_1P_2})+E_\alpha(\rho_{P_1P_3})\label{e1}
\end{align}
In  this way, we get the upper and lower bounds for $E_\alpha(\rho_{P_1|P_2P_3})$ when $\alpha\in[(\sqrt7-1)/2,(\sqrt{13}-1)/2], \alpha\neq1$. See Figure 1.
\begin{figure}[htpb]
\renewcommand{\figurename}{Fig.}
\centering
\includegraphics[width=6.5cm]{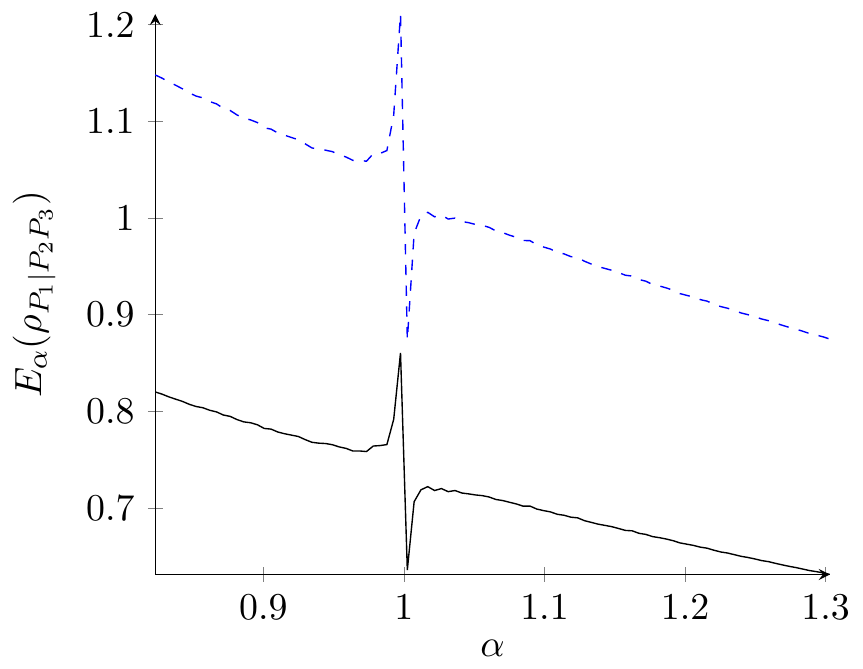}
\caption{{\small Solid  line is the function $\sqrt{E_\alpha^2(\rho_{P_1P_2})+E_\alpha^2(\rho_{P_1P_3})}$. Dashed blue line is the function $E_\alpha(\rho_{P_1P_2})+E_\alpha(\rho_{P_1P_3})$. }}
\label{Fig.1}
\end{figure}

\subsection{Upper bound for generalized W-class states using R\'{e}nyi entropy}\label{B}
The concurrence is related to the linear entropy of a state \cite{ym26},
\begin{align}
T(\rho)=1-Tr(\rho^{2})
\end{align}
Given a bipartite state $\rho_{AB}$, $T(\rho)$ has the property \cite{ym27}
\begin{align}\label{additive}
|T(\rho_A)-T(\rho_B)|\leq T(\rho_{AB})\leq T(\rho_A)+T(\rho_B).
\end{align}

Assume $\ket{\psi}_{PQR_1R_2\cdots R_{k-2}}$ is a GW state, from the definition of pure state concurrence together with Eq.(\ref{additive}), we have
\begin{align}\label{con-additive1}
|C^2(\ket{\psi}_{P|QR_1R_2\cdots R_{k-2}})-C^2(\ket{\psi}_{Q|PR_1R_2\cdots R_{k-2}})|\leq C^2(\ket{\psi}_{PQ|R_1R_2\cdots R_{k-2}}),
\end{align}
\begin{align}\label{con-additive2}
 C^2(\ket{\psi}_{PQ|R_1R_2\cdots R_{k-2}})\leq C^2(\ket{\psi}_{P|QR_1R_2\cdots R_{k-2}})+C^2(\ket{\psi}_{Q|PR_1R_2\cdots R_{k-2}}).
\end{align}
\begin{theorem}\label{T3}
 	Assume $\ket{\psi}_{PQR_1R_2\cdots R_{k-2}}$ is a GW state, when $\alpha\in[(\sqrt7-1)/2,(\sqrt{13}-1)/2]$, we have
 	\begin{align}
 	E_\alpha(\ket{\psi}_{PQ|R_1R_2\cdots R_{k-2}})\leq 2E_\alpha(\rho_{PQ})+\sum_{i=1}^{k-2}E_\alpha(\rho_{PR_i})+\sum_{i=1}^{k-2}E_\alpha(\rho_{QR_i}).
 	\end{align}
 \end{theorem}
[Proof]
For simplicity, we denote $R=R_1R_2\cdots R_{k-2}$.

In Lemma 2 of Ref.\cite{WS2}, the authors show
\begin{align}\label{WS2}
 	f_\alpha(x^2+y^2)\leq f_\alpha(x^2)+f_\alpha(y^2).
 	\end{align}
with $\alpha\in[(\sqrt7-1)/2,(\sqrt{13}-1)/2]$ and $ 0\leq x, y\leq1, 0\leq x^2+y^2 \leq 1$.

Then we have
\begin{align}\label{37}
	E_\alpha(\ket{\psi}_{PQ|R})=&f_\alpha(C^2(\ket{\psi}_{PQ|R}))\nonumber\\
\leq & f_\alpha(C^2(\ket{\psi}_{P|QR})+C^2(\ket{\psi}_{Q|PR}))\nonumber\\
	\leq&f_\alpha\left(C^2(\ket{\psi}_{P|QR})\right)+f_\alpha\left(C^2(\ket{\psi}_{Q|PR})\right)\nonumber\\
	=& f_\alpha\left(C^2(\rho_{PQ})+C^2(\rho_{P|R})\right)+f_\alpha\left(C^2(\rho_{QP})+C^2(\rho_{Q|R})\right)\nonumber\\
\leq & 2E_\alpha(\rho_{PQ})+f_\alpha(C^2(\rho_{P|R}))+f_\alpha(C^2(\rho_{Q|R})),
\end{align}
where in the second inequality we use (\ref{con-additive2}) and the monotonically increasing property of $f_\alpha(x)$ in Lemma 2. The third inequality is due to (\ref{WS2}). The forth equality is due to Lemma 5. Using (\ref{WS2}) again, we get the last inequality.

Since $R=R_1R_2\cdots R_{k-2}$, then
\begin{align}\label{38}
 	f_\alpha(C^2(\rho_{P|R}))=f_\alpha\left(\sum_{i=1}^{k-2}C^2(\rho_{PR_i})\right)
 \leq \sum_{i=1}^{k-2}f_\alpha(C^2(\rho_{PR_i}))
 =\sum_{i=1}^{k-2}E_\alpha(\rho_{PR_i}).
 	\end{align}
where the first equality is due to Lemma 5 and the second inequality is from the iterative use of (\ref{WS2}).
Similarly, we get
\begin{align}\label{39}
 	f_\alpha(C^2(\rho_{Q|R}))=f_\alpha\left(\sum_{i=1}^{k-2}C^2(\rho_{QR_i})\right)
 \leq \sum_{i=1}^{k-2}f_\alpha(C^2(\rho_{QR_i}))
 =\sum_{i=1}^{k-2}E_\alpha(\rho_{QR_i}).
 	\end{align}

Finally, combining (\ref{37}), (\ref{38}) and (\ref{39}), we complete the proof.
\qed

\begin{theorem}\label{T4}
 	Assume $\rho_{A_{j_1}A_{j_2}\cdots A_{j_m}}$ is the reduced density matrix of a GW state $\ket{\psi_{A_1\cdots A_n}},$ here we denote $\{P_1,P_2,P_3\}$ is a partition of the set $\{A_{j_1},A_{j_2},\cdots,A_{j_m}\},$ when $\alpha\in[(\sqrt7-1)/2,(\sqrt{13}-1)/2]$, we have the following monogamy inequality, we have
 	\begin{align}
 	E_\alpha ^a(\rho_{P_1|P_2P_3})\leq E_\alpha ^a(\rho_{P_2|P_1P_3})+E_\alpha ^a(\rho_{P_3|P_1P_2}).
 	\end{align}
 \end{theorem}
[proof]

Assume $\{p_{i}, |\psi_{i}\rangle\}$ is the optimal decomposition for REoA of $\rho_{P_1|P_2P_3}$ such that
$E_\alpha ^a(\rho_{P_1|P_2P_3})=\sum_i p_i E_\alpha (\ket{\psi_i}_{P_1|P_2P_3})$.

Let $E(\rho) = 2\left(1-Tr(\rho^2)\right)$.
For each pure state $\ket{\psi_i}_{P_1|P_2P_3}$ in this optimal decomposition with $\rho^i_{P_2P_3}= Tr_{P_1}|\psi _i \rangle _{P_1P_2P_3} \langle \psi_i|$, $\rho^i_{P_2}=Tr_{P_1P_3}|\psi _i \rangle _{P_1P_2P_3} \langle \psi_i|$, $\rho^i_{P_3}=Tr_{P_1P_2}|\psi _i \rangle _{P_1P_2P_3} \langle \psi_i|$, we have
\begin{align}\label{ABC}
 E_\alpha (\ket{\psi_i}_{P_1|P_2P_3})=&f_\alpha(C^2(\ket{\psi_i}_{P_1|P_2P_3}))\nonumber\\
 =&f_\alpha(E(\rho^i_{P_2P_3}))\nonumber\\
 \leq& f_\alpha(E(\rho^i_{P_2})+E(\rho^i_{P_3}))\nonumber\\
 =& f_\alpha[C^2(\ket{\psi_i}_{P_2|P_1P_3})+C^2(\ket{\psi_i}_{P_3|P_1P_2}))]\nonumber\\
 \leq& f_\alpha[C^2(\ket{\psi_i}_{P_2|P_1P_3})]+f_\alpha[C^2(\ket{\psi_i}_{P_3|P_1P_2}))]\nonumber\\
 =& E_\alpha(\ket{\psi_i}_{P_2|P_1P_3})+E_\alpha(\ket{\psi_i}_{P_3|P_1P_2}).
\end{align}
where in the first inequality we use the subadditivity of concurrence \cite{ym26} and the monotonically increasing property of $f_\alpha(x)$ in Lemma 2. The second inequality is due to (\ref{WS2}).

Then we have
\begin{align}
E_\alpha ^a(\rho_{P_1|P_2P_3})=&\sum_i p_i E_\alpha (\ket{\psi_i}_{P_1|P_2P_3})\nonumber\\
\leq& \sum_i p_iE_\alpha(\ket{\psi_i}_{P_2|P_1P_3})+\sum_i p_iE_\alpha(\ket{\psi_i}_{P_3|P_1P_2})\nonumber\\
\leq& E_\alpha ^a(\rho_{P_2|P_1P_3})+E_\alpha ^a(\rho_{P_3|P_1P_2}).
\end{align}
where the first inequality is due to inequality~(\ref{ABC}) and the second inequality is due to the definition of REoA.
\qed

According to Theorem~\ref{Ralpha3}, we have the following corollary.

\begin{corollary}\label{T5}
 	Assume $\rho_{A_{j_1}A_{j_2}\cdots A_{j_m}}$ is the reduced density matrix of a GW state $\ket{\psi_{A_1\cdots A_n}},$ here we denote $\{P_1,P_2,P_3\}$ is a partition of the set $\{A_{j_1},A_{j_2},\cdots,A_{j_m}\},$ when $\alpha\in[(\sqrt7-1)/2,(\sqrt{13}-1)/2]$, we have the following monogamy inequality, we have
 	\begin{align}
 	E_\alpha(\rho_{P_1|P_2P_3})\leq E_\alpha(\rho_{P_2|P_1P_3})+E_\alpha(\rho_{P_3|P_1P_2}).
 	\end{align}
 \end{corollary}
By Corollary~\ref{T5} , we have the upper bound for R$\alpha$E.
\begin{corollary}
Assume $\rho_{A_{j_1}A_{j_2}\cdots A_{j_m}}$ is the reduced density matrix of a GW state $\ket{\psi_{A_1\cdots A_n}},$ and here we denote $\{P_1,P_2,Q_1,Q_2,\cdots,Q_k\}$ is a partition of the set $\{A_{j_1},A_{j_2},\cdots,A_{j_m}\},$ when $\alpha\in[(\sqrt7-1)/2,(\sqrt{13}-1)/2]$, we have the following monogamy inequality,
\begin{align}
E_\alpha(\rho_{P_1P_2|Q_1\cdots Q_k})\leq 2E_\alpha(\rho_{P_1P_2})+\sum_{i=1}^kE_\alpha(\rho_{P_1Q_i})+\sum_{i=1}^kE_\alpha(\rho_{P_2Q_i}).
\end{align}
\end{corollary}

As an example, we consider a 4-qubit generalized W-class state
\begin{align}
&\ket{\psi}_{A_1A_2A_3A_4}
=a_1\ket{1000}+a_2\ket{0100}+a_3\ket{0010}+a_4\ket{0001},
\end{align}
with $\sum_{i=1}^4a_i^2=1$.

We choose $P_1={A_1}, P_2=\{A_2,A_3\}, P_3={A_4}$, then $\ket{\psi}_{A_1A_2A_3A_4}$ can be rewritten as
\begin{align}
&\ket{\psi}_{P_1P_2P_3}
=a_1\ket{1}\otimes\ket{00}\otimes\ket{0}+\sqrt{a_2^2+a_3^2}\ket{0}\otimes\left(\frac{a_2}{\sqrt{a_2^2+a_3^2}}\ket{10}+\frac{a_3}{\sqrt{a_2^2+a_3^2}}\ket{01}\right)\otimes\ket{0}++a_4\ket{0}\otimes\ket{00}\otimes\ket{1},
\end{align}

 After calculation, we have
$C^2(\ket{\psi}_{P_1P_2|P_3})=2[1-Tr(\rho_{P_1P_2}^{2})]$ with $Tr(\rho_{P_1P_2}^{2})=a_1^4+2a_1^2(a_2^2+a_3^2)+(a_2^2+a_3^2)^2+a_4^4$, $C^2(\rho_{P_1P_2})=4a_1^2(a_2^2+a_3^2)$,$C^2(\rho_{P_1P_3})=4a_1^2a_4^2$, and $C^2(\rho_{P_2P_3})=4a_4^2(a_2^2+a_3^2)$.
Set $a_1=\frac{3}{4}, a_2=\frac{1}{2}, a_3=\frac{\sqrt{2}}{4}, a_4=\frac{1}{4}$ ,
we plot the relation$E_\alpha(\ket{\psi}_{P_1P_2|P_3})\leq 2E_\alpha(\rho_{P_1P_2})+E_\alpha(\rho_{P_1P_3})+E_\alpha(\rho_{P_2P_3})$ in Theorem~\ref{T3} with $\alpha\in[(\sqrt7-1)/2,(\sqrt{13}-1)/2], \alpha\neq1$ in Figure 2.

\begin{figure}[htpb]
\renewcommand{\figurename}{Fig.}
\centering
\includegraphics[width=6.5cm]{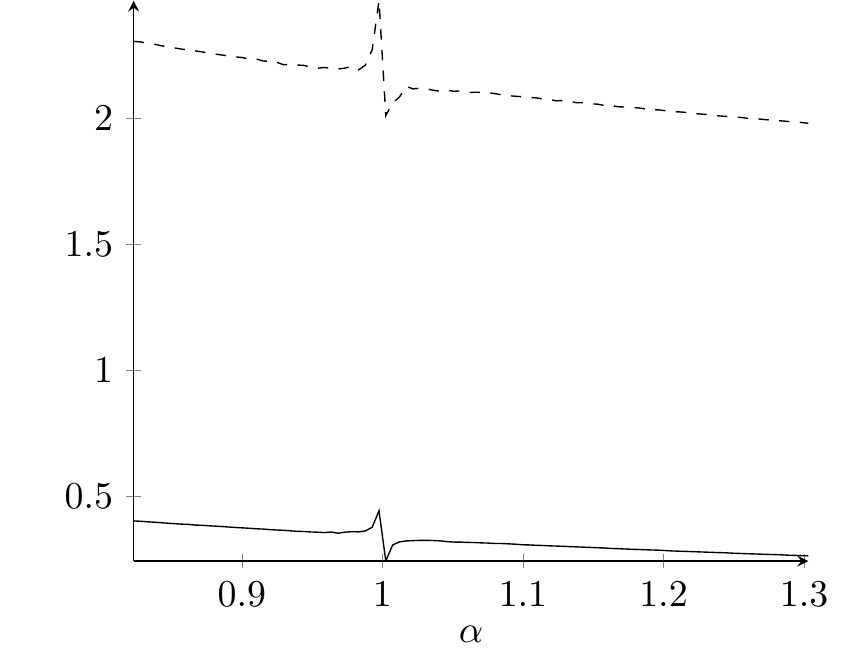}
\caption{{\small Solid  line is the function $E_\alpha(\ket{\psi}_{P_1P_2|P_3})$. Dashed line is the function $2E_\alpha(\rho_{P_1P_2})+E_\alpha(\rho_{P_1P_3})+E_\alpha(\rho_{P_2P_3})$, which is the upper bound for $E_\alpha(\ket{\psi}_{P_1P_2|P_3})$. }}
\label{Fig.2}
\end{figure}

\subsection{Tighter monogamy relations for generalized W-class state}\label{C}
If we set the partition $\{P_1,P_2,P_3\}$ is a subset of the set $\{A,B_1,B_2,...,B_{n-1}\}$, then the inequalities
(\ref{lemma5-1}) and (\ref{lemma5-2}) in Lemma 5 can be written as
\begin{align}\label{tighter1}
&C^2_{P_1|P_2P_3}
=C^2_{P_1P_2}+C^2_{P_1P_3},
\end{align}
\begin{align}\label{tighter2}
&C_{P_1P_2}
=C^a_{P_1P_2}.
\end{align}

\begin{theorem}\label{general1}
 	Assume $\ket{\psi}_{AB_1B_2\cdots B_{n-1}}$ is a GW state and set the partition $\{P_1,P_2,P_3\}$ is a subset of the set $\{A,B_1,B_2,...,B_{n-1}\}$, if $C^\alpha_{P_1P_3}\geq kC^\alpha_{P_1P_2}$, we have
 	\begin{align}\label{tighter3}
 	\left(C^a_{P_1|P_2P_3}\right)^\beta\geq h\left(C^a_{P_1P_3}\right)^\beta+\left(C^a_{P_1P_2}\right)^\beta.
 	\end{align}
with $\beta\in[0,\alpha], \alpha \geq 2, h=\frac{(1+k)^\frac{\beta}{\alpha}-1}{k^\frac{\beta}{\alpha}}, k \geq 1$.
 \end{theorem}
[Proof]
Since $C^\alpha_{P_1P_3}\geq kC^\alpha_{P_1P_2}$, then we have
\begin{align}
 	\left(C^a_{P_1|P_2P_3}\right)^\beta=&\left(C_{P_1|P_2P_3}\right)^\beta\nonumber\\
 \geq&\left(C^\alpha_{P_1P_2}+C^\alpha_{P_1P_3}\right)^\frac{\beta}{\alpha}\nonumber\\
 =&C_{P_1P_2}^\beta\left(1+\frac{C_{P_1P_3}^\alpha}{C_{P_1P_2}^\alpha}\right)^\frac{\beta}{\alpha}\nonumber\\
 \geq& C_{P_1P_2}^\beta\left[1+\frac{(1+k)^\frac{\beta}{\alpha}-1}{k^\frac{\beta}{\alpha}}\left(\frac{C_{P_1P_3}^\alpha}{C_{P_1P_2}^\alpha}\right)^\frac{\beta}{\alpha}\right]\nonumber\\
 =&C_{P_1P_2}^\beta+\frac{(1+k)^\frac{\beta}{\alpha}-1}{k^\frac{\beta}{\alpha}}C_{P_1P_3}^\beta\nonumber\\
 =&\left(C^a_{P_1P_3}\right)^\beta+\frac{(1+k)^\frac{\beta}{\alpha}-1}{k^\frac{\beta}{\alpha}}\left(C^a_{P_1P_2}\right)^\beta.
 	\end{align}
Here the first inequality is due to (\ref{tighter1}). The second inequality is obtained from Lemma 7 and the last equality is due to (\ref{tighter2}).

\qed

 Theorem \ref{general1} gives us a general monogamy inequality for the GW states using CoA which is tighter than the result in Ref.\cite{2020}. Next we present an example for Theorem \ref{general1}.

As an example, we consider a three-qubit generalized state
\begin{align}
&\ket{\psi}_{A_1A_2A_3}
=\frac{1}{6}\ket{100}+\frac{1}{6}\ket{010}+\frac{2}{\sqrt{6}}\ket{001}.
\end{align}
Then we have $C(\ket{\psi}_{A_1A_2A_3})=\frac{\sqrt{5}}{3}, C(\rho_{AB})=C^a(\rho_{AB})=\frac{1}{3}, C(\rho_{AC})=C^a(\rho_{AC})=\frac{2}{3}$. Choose $\alpha=2$, since $1\leq k \leq 4$, Set $k=2$, we can see that our result is better than the result in Ref.\cite{2020} from Figure.3.
\begin{figure}[htpb]
\renewcommand{\figurename}{Fig.}
\centering
\includegraphics[width=6.5cm]{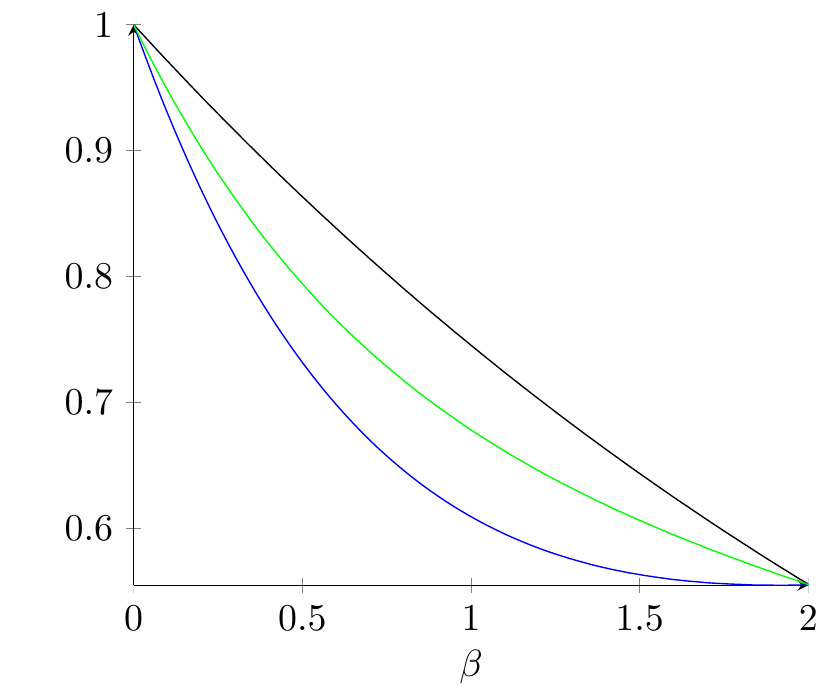}
\caption{{\small Black line is the function $C(\ket{\psi}_{A_1|A_2A_3})$. Blue line is the result in Ref.\cite{2020}. The green line is the lower bound from our result.}}
\label{Fig.3}
\end{figure}

Next we generalize the results to multipartite GW state. This results states all the powers of the GW state in terms of CoA under some restricted conditions.
\begin{theorem}\label{general2}
 Assume $\rho_{P_{1}\cdots P_{m}}$ is the reduced density matrix of a GW state $\ket{\psi}_{AB_1\cdots B_{n-1}},$
if $kC^\alpha_{P_1P_i}\leq C^\alpha_{P_1|P_{i+1}\cdots P_{m-1}}$ for $i=2, 3, \cdots, n$, and
$C^\alpha_{P_1P_j}\geq kC^\alpha_{P_1|P_{j+1}\cdots P_{m}}$ for $j=n+1,\cdots,m-1$,   then we have
\begin{align}\label{tighter4}
 	\left(C^a_{P_1|P_2...P_m}\right)^\beta\geq \sum_{i=2}^{n}h^{i-2}\left(C^a_{P_1P_i}\right)^\beta+h^{n}\sum_{i=n+1}^{m-1}\left(C^a_{P_1P_i}\right)^\beta+h^{n-1}\left(C^a_{P_1P_m}\right)^\beta.
 	\end{align}
with $\beta\in[0,\alpha], \alpha \geq 2, h=\frac{(1+k)^\frac{\beta}{\alpha}-1}{k^\frac{\beta}{\alpha}}, k \geq 1$.
\end{theorem}
[Proof]
Since $kC^\alpha_{P_1P_i}\leq C^\alpha_{P_1|P_{i+1}\cdots P_{m-1}}$ for $i=2, 3, \cdots, n$, then using Theorem~\ref{general1}, we have
\begin{align}\label{tighter4-1}
 	\left(C^a_{P_1|P_2...P_m}\right)^\beta
 \geq&\left(C^a_{P_1P_2}\right)^\beta+h\left(C^a_{P_1|P_3...P_m}\right)^\beta\nonumber\\
 \geq& \cdots\nonumber\\
 \geq& \sum_{i=2}^{n}h^{i-2}\left(C^a_{P_1P_i}\right)^\beta+h^{n-1}\left(C^a_{P_1|P_{n+1}...P_m}\right)^\beta.
 \end{align}

Since $C^\alpha_{P_1P_j}\geq kC^\alpha_{P_1|P_{j+1}\cdots P_{m}}$ for $j=n+1,\cdots,m-1$, using Theorem~\ref{general1} again, we have
\begin{align}\label{tighter4-2}
 	\left(C^a_{P_1|P_{n+1}...P_m}\right)^\beta
 \geq&h\left(C^a_{P_1P_{n+1}}\right)^\beta+\left(C^a_{P_1|P_{n+2}...P_m}\right)^\beta\nonumber\\
 \geq& \cdots\nonumber\\
 \geq& h\sum_{i=n+1}^{m-1}\left(C^a_{P_1P_i}\right)^\beta+\left(C^a_{P_1P_m}\right)^\beta.
 \end{align}

Combining (\ref{tighter4-1}) and (\ref{tighter4-2}), we have
\begin{align}\label{tighter4-3}
 	\left(C^a_{P_1|P_2...P_m}\right)^\beta
 \geq& \sum_{i=2}^{n}h^{i-2}\left(C^a_{P_1P_i}\right)^\beta+h^{n-1}\left(C^a_{P_1|P_{n+1}...P_m}\right)^\beta\nonumber\\
 \geq& \sum_{i=2}^{n}h^{i-2}\left(C^a_{P_1P_i}\right)^\beta+h^{n}\sum_{i=n+1}^{m-1}\left(C^a_{P_1P_i}\right)^\beta+h^{n-1}\left(C^a_{P_1P_m}\right)^\beta.
 \end{align}
\qed

CREN is equivalent to concurrence for any pure state with Schmidt rank two \cite{yym18}. So for any two-qubit mixed state $\rho_{AB}$,
\begin{align}
\mathcal{C}(\rho_{AB}) = \min_{\{p_{i}, |\psi_{i}\rangle\}}\sum_{i}p_{i}\mathcal{C}(|\psi_{i}\rangle)
=  \min_{\{p_{i}, |\psi_{i}\rangle\}}\sum_{i}p_{i}\mathcal{N}(|\psi_{i}\rangle)
=\widetilde{\mathcal{N}}(\rho_{AB}),
\end{align}
\begin{align}
\mathcal{C}_{a}(\rho_{AB}) = \max_{\{p_{i}, |\psi_{i}\rangle\}}\sum_{i}p_{i}\mathcal{C}(|\psi_{i}\rangle)
=  \max_{\{p_{i}, |\psi_{i}\rangle\}}\sum_{i}p_{i}\mathcal{N}(|\psi_{i}\rangle)
=\widetilde{\mathcal{N}}_{a}(\rho_{AB}),
\end{align}

For GW states, using the similar methods in Theorem ~\ref{general1} and Theorem ~\ref{general2}, we have the results for CREN.
\begin{theorem}\label{general13}
 	Assume $\ket{\psi}_{AB_1B_2\cdots B_{n-1}}$ is a GW state and set the partition $\{P_1,P_2,P_3\}$ is a subset of the set $\{A,B_1,B_2,...,B_{n-1}\}$, if $\widetilde{\mathcal{N}}^\alpha_{P_1P_3}\geq k\widetilde{\mathcal{N}}^\alpha_{P_1P_2}$, we have
 	\begin{align}
 	\left(\widetilde{\mathcal{N}}_{P_1|P_2P_3}\right)^\beta\geq h\left(\widetilde{\mathcal{N}}_{P_1P_3}\right)^\beta+\left(\widetilde{\mathcal{N}}_{P_1P_2}\right)^\beta.
 	\end{align}
with $\beta\in[0,\alpha], \alpha \geq 2, h=\frac{(1+k)^\frac{\beta}{\alpha}-1}{k^\frac{\beta}{\alpha}}, k \geq 1$.
 \end{theorem}
\begin{theorem}\label{general4}
 Assume $\rho_{P_{1}\cdots P_{m}}$ is the reduced density matrix of a GW state $\ket{\psi}_{AB_1\cdots B_{n-1}},$
if $k\widetilde{\mathcal{N}}^\alpha_{P_1P_i}\leq \widetilde{\mathcal{N}}^\alpha_{P_1|P_{i+1}\cdots P_{m-1}}$ for $i=2, 3, \cdots, n$, and
$\widetilde{\mathcal{N}}^\alpha_{P_1P_j}\geq k\widetilde{\mathcal{N}}^\alpha_{P_1|P_{j+1}\cdots P_{m}}$ for $j=n+1,\cdots,m-1$,   then we have
\begin{align}
 	\left(\widetilde{\mathcal{N}}_{P_1|P_2...P_m}\right)^\beta\geq \sum_{i=2}^{n}h^{i-2}\left(\widetilde{\mathcal{N}}_{P_1P_i}\right)^\beta+h^{n}\sum_{i=n+1}^{m-1}\left(\widetilde{\mathcal{N}}_{P_1P_i}\right)^\beta+h^{n-1}\left(C^a_{P_1P_m}\right)^\beta.
 	\end{align}
with $\beta\in[0,\alpha], \alpha \geq 2, h=\frac{(1+k)^\frac{\beta}{\alpha}-1}{k^\frac{\beta}{\alpha}}, k \geq 1$.
\end{theorem}
Finally, we show the results for R$\alpha$E.
\begin{theorem}\label{general5}
 	Assume $\ket{\psi}_{AB_1B_2\cdots B_{n-1}}$ is a GW state and set the partition $\{P_1,P_2,P_3\}$ is a subset of the set $\{A,B_1,B_2,...,B_{n-1}\}$, if $C^\mu_{P_1P_3}\geq kC^\mu_{P_1P_2}$, when $\alpha\geq(\sqrt 7  - 1)/2$, we have
 	\begin{align}
 	\left(E_\alpha(\rho_{P_1|P_2P_3})\right)^\beta\geq h\left(E_\alpha(\rho_{P_1P_3})\right)^\beta+\left(E_\alpha(\rho_{P_1P_2})\right)^\beta.
 	\end{align}
with $\beta\in[0,\mu], \mu \geq 2, h=\frac{(1+k)^\frac{\beta}{\mu}-1}{k^\frac{\beta}{\mu}}, k \geq 1$.
 \end{theorem}
[Proof]
Since
\begin{align}
 	\left(f_\alpha(x^2+y^2)\right)^\beta\geq& \left(f_\alpha(x^2)+f_\alpha(y^2)\right)^\beta\nonumber\\
 \geq& \left(f_\alpha(x^2)\right)^\beta+\frac{(1+k)^\frac{\beta}{\mu}-1}{k^\frac{\beta}{\mu}}\left(f_\alpha(y^2)\right)^\beta.
 	\end{align}
Here the first inequality is due to the convex property of $f_\alpha(x)$ for $\alpha\geq(\sqrt 7  - 1)/2$ in Lemma 3. The second inequality is obtained from a similar consideration in the proof of Theorem~~\ref{general1}.

Then we have
\begin{align}
 	\left(E_\alpha(\rho_{P_1|P_2P_3})\right)^\beta=&\left(f_\alpha(C^2(\rho_{P_1|P_2P_3}))\right)^\beta\nonumber\\
 =&\left(f_\alpha(C^2(\rho_{P_1P_2})+C^2(\rho_{P_1P_2}))\right)^\beta\nonumber\\
 \geq&\left(f_\alpha(C^2(\rho_{P_1P_2}))\right)^\beta+\frac{(1+k)^\frac{\beta}{\mu}-1}{k^\frac{\beta}{\mu}}\left(f_\alpha(C^2(\rho_{P_1P_3}))\right)^\beta\nonumber\\
 =& \left(E_\alpha(\rho_{P_1P_2})\right)^\beta+\frac{(1+k)^\frac{\beta}{\mu}-1}{k^\frac{\beta}{\mu}}\left(E_\alpha(\rho_{P_1P_3})\right)^\beta.
 	\end{align}
\qed

Taking the similar consideration of Theorem~\ref{general2} to generalize Theorem ~\ref{general5} , we have the following results.
\begin{theorem}
 Assume $\rho_{P_{1}\cdots P_{m}}$ is the reduced density matrix of a GW state $\ket{\psi}_{AB_1\cdots B_{n-1}},$
if $kC^\mu_{P_1P_i}\leq C^\mu_{P_1|P_{i+1}\cdots P_{m-1}}$ for $i=2, 3, \cdots, n$, and
$C^\mu_{P_1P_j}\geq kC^\mu_{P_1|P_{j+1}\cdots P_{m}}$ for $j=n+1,\cdots,m-1$, when $\alpha\geq(\sqrt 7  - 1)/2$,  then we have
\begin{align}
 	\left(E_\alpha(\rho_{P_1|P_2...P_m})\right)^\beta\geq \sum_{i=2}^{n}h^{i-2}\left(E_\alpha(\rho_{P_1P_i})\right)^\beta+h^{n}\sum_{i=n+1}^{m-1}\left(E_\alpha(\rho_{P_1P_i})\right)^\beta+h^{n-1}\left(E_\alpha(\rho_{P_1P_m})\right)^\beta.
 	\end{align}
with $\beta\in[0,\mu], \mu \geq 2, h=\frac{(1+k)^\frac{\beta}{\mu}-1}{k^\frac{\beta}{\mu}}, k \geq 1$.
\end{theorem}

Kim and Sanders in Ref.~\cite{KS} propose a class of mixed states
\begin{equation}\label{mix}
\rho_{A_1\cdots A_n} = p \ket{W_n^d}_{A_1\cdots A_n}\bra{W_n^d} + (1-p)\ket{0\cdots0}_{A_1...A_n}\bra{0\cdots0}.
\end{equation}
for $0\leq p \leq 1$.
They further prove this kind of states satisfy the monogamy relation for concurrence.
Since $\rho_{A_1\cdots A_n}$ is an operator of rank two, we can always have a purification of $\rho_{A_1\cdots A_n}$ such that
\begin{eqnarray}\label{purification}
\ket{\psi}_{A_1\cdots A_n A_{n+1}}&=&\sqrt{p}\ket{W_n^d}_{A_1\cdots A_n}\otimes\ket{0}_{A_{n+1}}\nonumber\\
&&+\sqrt{1-p}\ket{0\cdots 0}_{A_1\cdots
A_n}\otimes\ket{x}_{A_{n+1}},
\end{eqnarray}
with $\ket{x}_{A_{n+1}}=\sum_{1=i}^{d-1}a_{n+1i}\ket{i}_{A_{n+1}}$ is a $1$-qudit quantum state of $A_{n+1}$.
(\ref{purification}) can be rewritten as
\begin{eqnarray}
\ket{\psi}_{A_1\cdots A_{n+1}}= \sum_{i=1}^{d-1}&&[\sqrt{p}(a_{1i}\ket{i\cdots00}_{A_1\cdots A_{n+1}}+\cdots+a_{ni} \ket{0\cdots i0})_{A_1\cdots A_{n+1}}\nonumber\\
&&+ \sqrt{1-p}a_{n+1i}\ket{0\cdots 0i}_{A_1\cdots A_{n+1}}], \label{puri2}
\end{eqnarray}
It is an $(n+1)$-qudit W-class state. So we conclude that the above results in this subsection are valid for mixed states in (\ref{mix}).

\section{Application in quantum games}\label{4}
In this section we reconsider the problem of quantum for GW states considered in Ref.\cite{quantumgame,2020} using R\'{e}nyi entropy.

A two-player game $G=(A,B,X,Y,\pi,v)$ is played between a referee and two isolated players, Alice and Bob, who communicate only with the referee and not between themselves. $\pi$ is a probability distribution: $X\times Y \longrightarrow  [0,1]$; $v$ is a verification function: $X\times Y \times A\times B \longrightarrow [0,1]$. The referee chooses a question pair $(x, y)$ on the question alphabets $X\times Y$ according to some probability distribution $\pi$ , then he sends $x$ to Alice and $y$ to Bob. Next the two players give their answers $a$ and $b$ from the sets $A$ and $B$.
If $v(x,y,a,b)=1$ for the verification function, then they win.
The classical value of the game \[cv(G)= \sup_{a_x,b_y } \sum_{x,y,a,b}\pi(x,y) v(a, b, x, y)   \int_\Omega a_x(\omega)b_y(\omega)d \mathbb{P}(\omega)\]
is the maximum winning probability when two platers can use optimal deterministic strategies $\sum_{a}a_x(\omega)=\sum_{b}b_y(\omega)=1$ based on some classical correlation $\mathbb{P}(\omega)$.
The quantum value for a bipartite entangled state $\rho_{AB}$ of the game is \[qv(G)= \sup_{\rho,E_x^a,F_{y}^b} \sum_{x,y,a,b}\pi(x,y)  v(a, b, x, y) tr(\rho E_x^a\otimes F_y^b)\]
where the maximum takes overall the POVMs ${E_x^a}$ and ${F_y^b}$, $\sum_{a}E_x^a=1, \sum_{b}F_y^b=1$.
It is clear that for all games, $cv(G)\leq qv(G)$.

In Ref.\cite{quantumgame}, the authors assume Alice has a $d$-dimensional system $A$. She can share quantum or classical correlation with an arbitrary number of players $B_1,B_2,...,B_n$, simultaneously. The referee randomly selects a player $B_i$ and plays the game $G_i=(A,B_i,X_i,Y_i,\pi_i,v_i)$ with Alice and $B_i$. For $\{G_i\}_{1\le i\le n}$, they defined the average entangled value:
\begin{align}
Aqv(\{G_i\})=\sup_{\rho,E_x^a, F_{1,y}^b,\cdots, F_{n,y}^b} \frac{1}{n}\sum_{i=1}^n\sum_{a,b,x,y}\pi_i(x,y) v_i(a, b, x, y)tr(\rho^{AB_i} E_x^a\otimes F_{i,y}^b),
\end{align}
here  $E_x^a, F_{1,y}^b,\cdots, F_{n,y}^b$ are POVMs on $A, B_1,\cdots, B_n$ respectively and $\rho^{AB_1\cdots B_n}$ is a multipartite state with $|A|$ is at most $d$. Since the classical correlation used for different $G_i$ can be combined, then
the average classical value was given by
\begin{align}
&Acv(\{G_i\})= \frac{1}{n}\sum_{i=1}^n cv(G_i).
\end{align}

In Ref.\cite{2020}, the authors reconsider the bound of the difference between the quantum games and the classical games restricting to GW states using Tsallis $q$-entropy for $q\in(1,2]$. In the following, we get a new bound of the difference between the quantum games and the classical games restricting to GW states using R\'{e}nyi $\alpha$-entropy for $\alpha\geq1$.
We use the similar method which has been used in Ref.\cite{quantumgame,2020}.
Let $G=(A,B,X,Y,\pi,v)$ be a quantum game. For fixed axillary systems $A,B$ and POVMs $E_x^a,F_y^b$, the value function becomes a positive linear function
\[lin_G(\rho_{AB})=\sum_{x,y,a,b}\pi(x,y)  v(a, b, x, y) tr(\rho E_x^a\otimes F_y^b).\]

Note that $lin_G$ is of norm at most $1$, then for a separable $\sigma_{AB}$ and an arbitrary $\rho_{AB}$,
\begin{align}
lin_G(\rho_{AB}) \leq lin_G(\rho_{AB} - \sigma_{AB})  + lin_G(\sigma_{AB})
\leq  \| \rho_{AB} - \sigma_{AB} \|_1 + cv({G}) .
\end{align}

For a bipartite pure state $\ket{\psi}_{AB}=\sum_{i=0}^{d-1}\sqrt{\lambda_i}\ket{ii},$ then we show there exists a separable state $\ket{\sigma}_{AB}$ such that
\begin{align}\label{pure}
\| \psi_{AB} - \sigma_{AB} \|_1 \leq 2\sqrt{2E_\alpha(\rho_{AB})}.
\end{align}
for $\alpha \geq 1$.
First we select $\sigma_{AB}=\ket{00}$ then we compute the trace norm $\| \ket{\psi}_{AB}\bra{\psi} - \ket{00}\bra{00} \|_1=2\sqrt{1-\lambda_0}$.
Then we show $2\sqrt{1-\lambda_0}\leq 2\sqrt{\frac{-2\log\sum_{i=0}^{d-1}\lambda_{i}^\alpha}{\alpha-1}}$.

Since when $\alpha \geq 1$, $\lambda\in[0,1]$, $\sum_{i=0}^{d-1}\lambda_{i}^\alpha \geq \lambda_{0}^\alpha+(1-\lambda_{0})^\alpha$,
then it is enough for us to show $2\sqrt{1-\lambda_0}\leq 2\sqrt{\frac{-2\log[\lambda_{0}^\alpha+(1-\lambda_{0})^\alpha]}{\alpha-1}}$.

Let $f_\alpha(\lambda_{0})=-2\log[\lambda_{0}^\alpha+(1-\lambda_{0})^\alpha]-(1-\lambda_{0})(\alpha-1)$, we need to show $f_\alpha(\lambda_{0})\geq 0$.
Since
\begin{align}\label{der}
f'_\alpha(\lambda_{0})=\frac{-2\alpha\lambda_{0}^{\alpha-1}+2\alpha(1-\lambda_0)^{\alpha-1}}{[\lambda_{0}^\alpha+(1-\lambda_{0})^\alpha]ln2}+(\alpha-1),
\end{align}
after analysis, we find (\ref{der}) has only one zero $\epsilon\in[0,1]$. When $\lambda_0\in[0,\epsilon],$ $f_\alpha(\lambda_{0})$ is monotonically increasing while monotonically decreasing when $\lambda_0\in[\epsilon,1]$.
Note that $\lambda_0\geq \frac{1}{d}$, then it is enough to show $f_\alpha(0)\geq 0$ and $f_\alpha(\frac{1}{d})\geq 0$.

$f_\alpha(0)\geq 0$ is clear. After computation, we get $f_\alpha(\frac{1}{d})=-\log[1+(d-1)^\alpha]+\alpha\log d-\frac{(d-1)(\alpha-1)}{d}$. It is easy to get $f_\alpha(\frac{1}{d})\geq0$ for any pure state with Schmidt rank equal or less than two when $\alpha\geq1$.

when $\rho$ is a mixed state, assume $\{p_{i}, |\psi_{i}\rangle\}$ is the optimal decomposition of $\rho$ in term of $E_\alpha(\rho_{AB})$, then
\begin{align}\label{mixed}
\| \rho_{AB} - \sigma_{AB} \|_1=&  \| \sum_{i} {p_i | \psi _i \rangle _{AB} \langle \psi _i |} - \sum_{i} {p_i | \theta _i \rangle _{AB} \langle \theta _i |} \|_1\nonumber\\
\leq &\sum_{i} {p_i}\| | \psi _i \rangle _{AB} \langle \psi _i | -  | \theta _i \rangle _{AB} \langle \theta _i | \|_1\nonumber\\
\leq &2\sqrt{2}\sum_{i} {\sqrt{p_i}}\sqrt{p_iE_\alpha(| \psi _i \rangle _{AB})}\nonumber\\
\leq&2\sqrt{2}\sqrt{E_\alpha(\rho_{AB})}.
\end{align}
Here we use the subadditivity of the 1-norm. The second inequality is due to (\ref{pure}) and the last inequality is due to the definition of REoA and Theorem~\ref{Ralpha3}.

By monogamy inequality in Theorem~\ref{monogamy1}, we have
\begin{align}\label{xiaoqualpha}
\sum_{i=1}^{n}E^2_\alpha(\rho_{AB_i})\leq E^2_\alpha(\rho_{A|B_1...B_n})\leq \left(\frac{\log d^{1-\alpha}}{1-\alpha}\right)^2=\left(\log d\right)^2.
\end{align}

Thus we have
\begin{align}
Aqv(G)\leq& \frac{2\sqrt{2}}{n}\sum_{i=1}^{n}\sqrt{E_\alpha(\rho_{A|B_1...B_n})}+Acv(G)\nonumber\\
\leq& \frac{2\sqrt{2}}{n^\frac{1}{4}}\sqrt{E_\alpha(\rho_{AB_i})}+Acv(G)\nonumber\\
\leq& \frac{2\sqrt{2}}{n^\frac{1}{4}}\left(\log d\right)^\frac{1}{2}+Acv(G),
\end{align}
where we use (\ref{mixed}) to get the first inequality. The second inequality is obtained from H{\"o}lder's inequality. The last inequality is due to (\ref{xiaoqualpha}).

So we have
 \begin{align}
Aqv(G)-Acv(G)
\leq& \frac{2\sqrt{2}}{n^\frac{1}{4}}\left(\log d\right)^\frac{1}{2},
\end{align}

We find the  bound of the difference between the quantum games and the classical games restricting to GW states using R\'{e}nyi $\alpha$-entropy for $\alpha\geq1$ is independent of $\alpha$. When $d=2$, the bound is the same as the bound obtained by Tsallis $q$-entropy for $q=2$ in Ref. \cite{2020} .

Compared the result in Ref. \cite{quantumgame}:
\begin{align}
Aqv(G)-Acv(G)
\leq& \frac{3.1}{n^\frac{1}{4}}d\left(\log d\right)^\frac{1}{4},
\end{align}
Our bound is tighter due to $d\geq \left(\log d\right)^\frac{1}{4}.$

\section{Conclusion}\label{5}
In this paper, we have investigated the general monogamy inequalities for the GW states using R$\alpha$E. First, we have shown an analytical formula of R$\alpha$E and REoA for a reduced density matrix of GW states. According to the analytical formula, we have presented a monogamy inequality in terms of the $\mu$-th power of R$\alpha$E for density matrices of GW states when $\mu \geq2$, and a polygamy inequality in terms of the $\mu$-th power of REoA for density matrices of GW states when $0 < \mu \leq1$. We also present the upper bounds for the GW states using  R$\alpha$E. Corresponding examples are also given . Then we have provided tighter monogamy relations in terms of concurrence and CREN. By the relation between R$\alpha$E and concurrence, we also obtain the general monogamy relations for R$\alpha$E. They are all also valid for a class of mixed states. Finally, we apply the monogamy relations to quantum games when restricting to the GW states. When $d=2$, our result is the same as the result obtained by Tsallis $q$-entropy for $q=2$ in Ref. \cite{2020}. Our result is also tighter than the result in Ref. \cite{quantumgame}. Moreover, our results in this paper will provide a reference for general monogamy and polygamy relations in multipartite higher dimensional quantum systems.

\bigskip
\noindent{\bf Acknowledgments}\, \, This work is supported by the NSFC 11571119 and Chinese Scholarship Council.

\end{document}